\documentclass[onecolumn,showpacs,prd]{revtex4}
\usepackage{epsfig}
\usepackage{bm}
\usepackage{amsfonts}
\usepackage{amssymb,amsmath}
\usepackage{enumerate}
\usepackage{multirow}
\usepackage{epsfig}
\usepackage{graphicx}
\usepackage{bm}
\usepackage{hyperref}

\begin{document}

\title{Non-minimally coupled quintessence DE model with a cubic galileon term --A Dynamical System Analysis}
\author{Somnath Bhattacharya}
\email{somnath94347@gmail.com}
\affiliation{Department of Physics, Barasat Govt. College, 10 K. N. C. Road, Barasat, Kolkata 700 124, India}
\author{Pradip Mukherjee}
\email{mukhpradip@gmail.com}
\affiliation{Department of Physics, Barasat Govt. College, 10 K. N. C. Road, Barasat, Kolkata 700 124, India}
\altaffiliation{Visiting Associate, Inter University Centre for Astronomy and Astrophysics, Pune, India}
\author{Anirban Saha}
\email{anirban@iucaa.ernet.in}
\affiliation{Department of Physics, West Bengal State University, Barasat, North 24 Paraganas, West Bengal, India}
\altaffiliation{Visiting Associate, Inter University Centre for Astronomy and Astrophysics, Pune, India}
\author{Amit Singha Roy}
\email{singharoyamit@gmail.com}
\affiliation{Department of Physics, Barasat Govt. College, 10 K. N. C. Road, Barasat, Kolkata 700 124, India}

\begin{abstract}
We consider a scalar field which is generallly non-minimally coupled to gravity and has a characteristic cubic Galilean-like term in the kinetic part of the action, in presence of a generic self-interaction as a candidate Dark Energy model. The system is dynamically analyzed and novel fixed points with perturbative stability are demonstrated. Evolution of the system is numerically studied near a novel fixed point which owes its existance to the Galileon character of the model. It turns out that demanding the stability of this novel fixed points puts strong restriction on the allowed non-minimal coupling and the choice of the self-interaction. The evolutions of the system is charted out on a $r-s$ diagram. The evolution of the equation of state parameter is studied which shows that our model predicts accelerated universe throughout and the phantom limit is only approached closely but never crossed. Our result thus extends the findings of of \cite{Cubic_Galileon_NMC} for more general NMC than linear and quadratic couplings. 
\end{abstract}

\maketitle

\section{Introduction}
Recent cosmological observations \cite{obe1, obe2, obe3, obe4, obe5,obe6, obe7, obe8, obe9, obe10, obe11, obe12, obe13, obe14, obe15, obe16, obe17, obe18, obe19} indicate late-time acceleration of the observable universe, a phenomenon that could not be explained by standard cosmology. Various theoretical attempts have been undertaken to confront this observational fact. The late time acceleration requires an equation of state (EOS) parameter $\omega < -\frac{1}{3}$ which can not be realized by ordinary matter or radiation. Thus one has to introduce either a cosmological constant (CC) in the field equation leading to the $\Lambda$CDM model with $ \omega =-1$ or an exotic dark energy (DE) fluid \cite{DE, Paddy_1, DE_Review_Sami} described by a scalar field with a suitable dynamics of its own. Though most observational evidance favours the former, there exist certain conceptual difficulties with it, e.g., the fine tuning and coincidence problems owing to the absence of dynamics of the CC. So we look at the dynamical alterntive, i.e. scalar field DE model. If this scalar field has a canonical kinetic energy term then $\omega$ is bounded within the range $-1 < \omega <-\frac{1}{3}$, this is called quintessence scenario. However in the past observational data \cite{Komatsu, obs_data_Phamtom_crossing} have allowed $\omega$ to cross the deSitter limit $ \omega =-1$. The recent Planck observations \cite{Planck_1} laso have produced various combined data sets on the EoS parameter some of which (e.g., $95 \%$; Planck+WMAP+SNLS and $95 \%$; Planck+WMAP+${\rm H}_{0}$) prefer certain phantom evolution of the DE fluid with $\omega < -1$.

To realize such an $\omega$ theoretically one ordinarily requires to change the sign of the kinetic energy term, thus invoking the phantom model \cite{Phantom_model_1, Phantom_model_2}, which is Quintessence with the wrong sign. But of course the phantom models is plagued with various instabilites \cite{Quantum_Instability_Phantom_1, Quantum_Instability_Phantom_2}. 
On the other hand a simple deviation from General Relativity (GR) can be useful in this context. This involves coupling of the scalar fields with curvature in addition to the minimal coupling as par GR and therefore called non-minimal coupling (NMC), which appeared much earlier in the literature in Brans-Dicke theory \cite{nmo2}. There are many theoretical arguments suggesting that the NMC should be considered. It comes from quantum corrections \cite{bd}, renormalization of classical theory \cite{cc}, in the string theoretic context \cite{sgc} or in the Scalar-tensor theories (STT) \cite{nmo2, nmo3, nmo4, nmo5, nmo6, nmo7, nmo8, nmo9, nmo10, nmo11}. In conjunction with the DE scalar field this artifice has been utilized in many recent studies in the literature \cite{NM_quintescence_1, NM_quintescence_2, NM_quintescence_3, ref4, ref5}. Specifically the NMC is found to influence the EOS parameter $\omega$ in comparison with the corresponding model with minimal coupling \cite{NMC_quintescence_Phantom_crossing_1, NMC_quintescence_Phantom_crossing_2} . Thus the whole gamut of the scalar field dark energy models is widened with new possibilities. These possibilities have been explored with different scalar field dynamics. Mention should be made in this context of \cite{ref5} where dynamical system analysis is used in a non minimally coupled quintissence model to identify the stable fixed points. It is found that such a model can exhibit phantom crossing. 

In recent times a lot of interest has been focussed on the DE models with Galileon scalar field \cite{Galilean_1, ref3, Galilean_2, Galilean_3} due to the absence of negative energy instability and curvature singularity. The Galileon model features an effective scalar field $\pi$ which arises in the decoupling limit of the DGP model \cite{DGP}. In the original Galileon scalar field cosmology \cite{Galilean_1}, the Galilean symmetry $\partial_{\mu} \pi \to \partial_{\mu} \pi + b_{\mu}$ is essentially broken when gravity is introduced. However, a covariant formulation of the Galileon model  has been obtained in \cite{Galileon_2nd_order_eqm_1} where the shift symmetry $\pi \to \pi + c $ is preserved and the Galilean symmetry is said to be softly broken. The model gives rise to late time acceleration of the universe \cite{AASen11} and is consistent with the solar system tests of gravity through the Vainshtein mechanism \cite{AASen12} which works due to the presence of certain non-linear kinetic energy terms like $\Box{\pi} (\nabla \pi)^{2}$ in the Lagrangian apart from a term linear in $\pi$ and the usual canonical kinetic term. In spite of the presence of such higher derivative term in the Lagrangian the equation of motion comes out to be second order \cite{Galileon_2nd_order_eqm_1, Galileon_2nd_order_eqm_2, Galileon_2nd_order_eqm_3}, thus avoiding the Ostrogradsky ghosts. 

Interestingly, Galileon models minimally coupled to gravity exhibits phantom phase \cite{Galileon_Phantom_M, Pakis}. This feature is sheared by Galileon scalar non-minimally coupled to gravity \cite{Galileon_Phantom_NM} as well. However, in \cite{Cubic_Galileon_NMC} it has been shown that no phantom behaviour arises in the late-time evolution for linearly or quadratically (non-minimally) coupled cubic-Galileon cosmology with a linear potential. An interesting question may be posed in this context, namely, {\it wheather a phantom phase appears or not in a cubic Galileon model} \footnote{Although it has been shown that the full Galileon field Lagrangian can have another two non-linear kinetic terms involving higher derivatives, as the cubic Galileon term contains the essential Galileon character, we, like many others, choose to work with that.} {\it with a more general NMC and an arbitrary power-law potential.} In the present paper we address this specific question using a through dynamical system analysis \cite{Wainwright, Coley} of the corresponding closed analytic equations describing the said model. 

At this point we must clarify that the dark energy (DE) scalar considered in this paper is not a scalar under the Galilean shift symmetry $\pi \to \pi + c $ which is broken by any potential more general than $V\left(\pi \right) = M^{3} \pi$. Thus our model can not be called a Galileon scalar model in the strict sence. However, we chose to call the scalar in our model, a Galileon-like scalar because the non-linear kinetic energy term in our scalar field action is reminiscent of the same. Thus our model may be called a {\it{non minimally coupled Galileon-like scalar field model}}. Alternatively, our model can also be called a non-minimally coupled quintessence model generalized by including a cubic Galileon-like kinetic term.

  In \cite{AASen} it has been shown that such phenomenological Galileon-like DE scalars are favoured by the observational data over Quintessence  scalars, though with a minimal coupling to gravity. Generalizing the term linear in the Galileon field by some power law potential allows us to accommodate both quintessence and the Galileon type of scalar fields at one go by proper choice of the coupling parameters. This enables us to compare our result with existing results in the literature \cite{Sami_NMC} in the appropriate limit. Apart from this convenience, there exists another reason behind our choice of such a potential which is far more binding. A crucially important result appeared in \cite{solar_system_rule_out} where it has been shown that a non-minimally coupled DE scalar field theory with non-standard/non-linear kinetic terms and Galilean shift-symmetry (like the Galileon field) is ruled out due to too large a variation in the effective Newton's constant. Therefore we do not want the non-minimally coupled DE scalar $\pi$ to be a scalar under the Galilean shift symmetry and we achieve this by keeping the potential generic initially. We will see in the course of the paper that as the dynamical system analysis unfolds this generic potential eventually gets restricted to the power-law form $V\left(\pi \right) = V_{0} \pi^{-c}$, if we demand the autonomous equations describing the evolution of the model to form a closed system. Further, if we demand the non-trivial fixed point owing to the non-linear(cubic) kinetic term (referred as Galilean-like kinetic term in the paper) be stable, a upperbound on the power law $c < -1$ appears. It was reassuring to see that demanding a stable fixed point owing to the non-linear kinetic term pushes the model away from the very symmetry which is in tension with the solar system constraints \cite{solar_system_rule_out}. As it turns out this specific fixed point also depends critically on the nature of NMC considered and demanding its stability puts a restriction on the allowed NMC as well.

Once we identify this particulr fixed point we will perform a through numerical analysis to study the behaviour of the system in the corresponding asymptotic limit. We visualize the evolution of the system around this fixed point using the $r-s$ diagram \cite{vsahni} technique. 
Remembering that NMC has the physical appeal of modifying the energy momentum tensor and thus the EOS parameter, we plot the time evolution of the EoS parameter to check for any possible appreance of the phantom phase.

The organization of the present paper is as follows. In the next section the action of our model is given and assuming a Friedmann--Lemaitre--Robertson--Walker (FLRW) space-time, the equations of motion are obtained both for the scalar field and gravity. The energy momentum tensor is identified. The equation of state parameter is written down in terms of the geometric quantities in the FLRW background. In Section III the autonomous equations are set-up. The closure of the set of autonomous variables is actuated by the choice of a power law potential and the power law non-minimal coupling. Fixed points of the model are computed and we identify the subset of fixed points which owe their existance to the presence of the characteristic Galileon interaction. In section IV we focus on these novel fixed points and investigate their perturbative stability. 
The system is then numerically integrated near the stable fixed point and its various interesting features are discussed. We studied the evolution of the system on a $r-s$ plane and computed the late-time behaviour of the model by tracking the time-evolution of the equation of state parameter. We conclude in section V.

\section{The model}
\noindent Our model consists of a DE scalar field non-minimally coupled to gravity in presence of dust matter and insignificant radiation. The action is
\begin{equation}
S=\int{{{d^4}x}{\sqrt{-g}}}\left[\frac{{M_{pl}}^2}{2}\left(1-{\frac{{\xi}{B(\pi)}}{{M_{pl}}^2}}\right)R + {\cal{L}}_{\pi}  \right]+{S_{m}} 
\label{action} 
\end{equation}
where ${{M_{pl}}^2}=\frac{1}{8{\pi}G}$ is reduced Plank mass and $\xi$ is a dimensionless non-minimal coupling parameter. $B(\pi)$ is a differentiable function of the scalar field $\pi$.  
We denote the determinant of $g_{\mu\nu}$ by $g$ and the dust matter action by $S_{m}$ whereas the DE fluid is represented by ${\cal{L}}_{\pi} $. To ensure that the DE scalar has specific Galileon-like character (as discussed in the introduction), we retain upto the first non-trivial term of the standard Galileon Lagrangian in ${\cal{L}}_{\pi} $ and absorbe the tadpole term $\pi$ in a generic self-interaction $V(\pi)$
\begin{eqnarray}
{\cal{L}}_{\pi} = - {\frac{1}{2}}{({\nabla{\pi}})^2}-\frac{\alpha}{2{{M_{pl}}^3}}{({\nabla{\pi}})^2}(\Box{\pi})-V(\pi)
\label{L_Galilean}
\end{eqnarray}
The parameter $\alpha$ signifies the strength of the Galileon-like character of the system and also srves to identify its effect on the system dynamics.
\noindent We assume a spatially flat homogeneous isotropic FLRW universe, 
\begin{equation}
{ds}^2=-{dt}^2+{{a(t)}^2}\left[{{{dx}_1}^2+{{dx}_2}^2+{{dx}_3}^2}\right]\label{flrw}
\end{equation}
in the present work. The arbitrary scale factor $a(t)$ at present epoch is normalized to unity. In the following we first work out the equations of motion of various fields and also the equations of state (EoS) of the DE fluid.
\subsection{Equations of motion} 
The pair of Friedmann equations  are obtained by varying the action (\ref{action}) with respect to the metric tensor $g_{\mu\nu}$. It has the following form,
\begin{eqnarray}
3{{M_{pl}}^2}{H^2} &=& \left(1-{\frac{{\xi}{B(\pi)}}{{M_{pl}}^2}}\right)^{-1} \left[{{\rho}_m}+\frac{{\dot\pi}^2}{2}\left(1-\frac{6{\alpha}H{\dot\pi}}{{M_{pl}}^3}\right)+V(\pi)+3{\xi}H{\dot\pi}{B'({\pi})}\right]
\label{f1} \\ 
R={\frac{1}{{M_{pl}}^2}}&&\left[-{{\dot\pi}^2} + 4V(\pi)+3{\xi}\left({3H{\dot\pi}{B'({\pi})}}+{{\ddot{\pi}}{B'({\pi})}}+{{\frac{1}{3}}RB({\pi})}+{{{\dot\pi}^2}{B''({\pi})}}\right)+{\rho_m}\right. \nonumber\\
&& -\left. \left(\frac{3{\alpha}{{\dot\pi}^2}{\ddot\pi}}{{M_{pl}}^3}
+ \frac{3{\alpha}H{{\dot\pi}^3}}{{M_{pl}}^3}\right)\right]
\label{f2}
\end{eqnarray}
Here $R$ is the Ricci scalar and dot denotes derivative with respect to the comoving time. In FLRW space-time (\ref{flrw}), it is expressed by
\begin{eqnarray}
R=6(2{H^2}+\dot{H})
\label{R_scalar} 
\end{eqnarray}
where $H\left( t \right) = \frac{\dot{a}}{a\left( t \right) }$. In (\ref{f1}) and (\ref{f2}) $\rho_m$ is the energy density of the ordinary matter, read off from the matter energy momentum tensor (EMT) $T_{\left(m\right) \mu \nu}$ obtained from varying the matter action $S_{m}$.


Similarly, the equation of motion for the DE scalar is obtained as,
\begin{equation}
{\ddot{\pi}}+3H{\dot\pi}+V{'}({\pi})+{\frac{1}{2}}{\xi}R{B'({\pi})}-{\frac{3{\alpha}{\dot\pi}}{{M_{pl}}^3}}\left({\dot\pi}{\dot{H}}+2H{\ddot\pi}+3{H^2}{\dot\pi}\right)=0\label{eom}
\end{equation}
Note that due to the presence of second derivatives of $\pi$  in (\ref{L_Galilean}), particular care must be taken to calculate the covariant derivatives. Interestingly, in spite of the Lagrangian being of higher derivative type the equation of motion for $\pi$ is second order.This is a well known characteristic of the Galileon field. 

For non-minimally coupled DE scalar fields the EMT is to be defined appropriately. This EMT is used to determine the pressure and energy density of the DE fluid which in turns fixes its EoS parameter. The importance of this EoS parameter in analyzing the cosmological model can hardly be overemphasized. In the next subsection we therefore workout the same for our model.
\subsection{The equation of state parameter}
The EoS parameter is defined as,
\begin{equation}
\omega_{\pi}=\frac{P_{\pi}}{\rho_{\pi}}\label{defnomeg}
\end{equation} 
where both $P_{\pi}$ and $\rho_{\pi}$ are obtained from the DE EMT $T_{\left( \pi \right)\mu\nu}$ as
\begin{equation}
\rho_{\pi}=T_{\left( \pi \right)0}{}^{0} \qquad {\rm{and}} \qquad P_{\pi}=T_{\left( \pi \right) i}{}^{i}
\end{equation}
in analogy with the matter EMT. For minimally coupled theories the EMT is entirely contributed by the non gravitational part of the action. In case of non-minimal coupling 
one again varies the action(\ref{action}) with respect to the metric $g_{\mu\nu}$ to obtain the field equations of gravity and it is cast in the standard form of Einstein's field equation, $G_{\mu\nu}=8{\pi}G\left(T_{\left(m\right) \mu\nu} + T_{\left( \pi \right)\mu\nu} \right)$ to read off the DE EMT from the r.h.s. Following this prescription the DE EMT for the present model is obtained as, 
\begin{eqnarray}
T_{\mu\nu} &=& {\theta_{\mu\nu}}(minimal)+ {\theta_{\mu\nu}}(Huggins)\label{eosp}
\end{eqnarray}
from (\ref{action}).
where apart from the usual contribution owing to the minimally coupled part of the scalar field
\begin{eqnarray}
{\theta_{\mu\nu}}(minimal)=&&\left[({\partial_\mu}\pi)({\partial_\nu}\pi)+6\sqrt{6}\alpha({\partial_\mu}{\partial_\nu}\pi-{\Gamma^{\gamma}_{\mu\nu}}{\partial_\gamma}\pi){g^{\alpha\beta}}({\partial_\alpha}\pi)({\partial_\beta}\pi)\right.\nonumber\\
&&\left.+6\sqrt{6}\alpha({\partial_\mu}\pi)({\partial_\nu}\pi){g^{\alpha\beta}}({\partial_\alpha}{\partial_\beta}\pi-{\Gamma^{\gamma}_{\alpha\beta}}{\partial_\gamma}\pi)-\frac{1}{2}{g_{\mu\nu}}{g^{\lambda\sigma}}{({\partial_\lambda}\pi)}{({\partial_\sigma}\pi)}\right.\nonumber\\
&&\left.-3\sqrt{6}\alpha{g_{\mu\nu}}{g^{\alpha\beta}}({\partial_\alpha}{\partial_\beta}\pi-{\Gamma^{\gamma}_{\alpha\beta}}{\partial_\gamma}\pi){g^{\lambda\sigma}}({\partial_\lambda}\pi)({\partial_\sigma}\pi)-{g_{\mu\nu}V(\pi)}\right.\nonumber\\
&&\left.+6\sqrt{6}\alpha({\nabla_\mu}({g^{\alpha\beta}}{{\partial_\alpha}\pi}{{\partial_\beta}\pi}{{\partial_\nu}\pi})-\frac{1}{2}{\nabla_\sigma}({g_{\mu\nu}}{g^{\alpha\beta}}{{\partial_\alpha}\pi}{{\partial_\beta}\pi}{{\partial^\sigma}\pi}))\right]\label{nonminimal}
\end{eqnarray} 
the it contains another term, known as the the Huggin's term $ {\theta_{\mu\nu}}({\rm Huggins})$ \cite{Huggins1}, \cite{Huggins2}
\begin{eqnarray}
{\theta_{\mu\nu}}(Huggins)&=&+\left[-\frac{1}{6(1-6{\xi}B(\pi))}\left\{{g_{\mu\nu}}{g^{\alpha\tau}}{\nabla_\alpha}{\nabla_\tau}(1-6{\xi}B(\pi))-{\nabla_\mu}{\nabla_\nu}(1-6{\xi}B(\pi))\right\}\right.\nonumber\\
&&\left.+\frac{6{\xi}B(\pi)}{(1-6{\xi}B(\pi))}\left\{({\partial_\mu}\pi)({\partial_\nu}\pi)+6\sqrt{6}\alpha({\partial_\mu}{\partial_\nu}\pi-{\Gamma^{\gamma}_{\mu\nu}}{\partial_\gamma}\pi){g^{\alpha\beta}}({\partial_\alpha}\pi)({\partial_\beta}\pi)\right.\right.\nonumber\\
&&\left.\left.+6\sqrt{6}\alpha({\partial_\mu}\pi)({\partial_\nu}\pi){g^{\alpha\beta}}({\partial_\alpha}{\partial_\beta}\pi-{\Gamma^{\gamma}_{\alpha\beta}}{\partial_\gamma}\pi)-\frac{1}{2}{g_{\mu\nu}}{g^{\lambda\sigma}}{({\partial_\lambda}\pi)}{({\partial_\sigma}\pi)}\right.\right.\nonumber\\
&&\left.\left.-3\sqrt{6}\alpha{g_{\mu\nu}}{g^{\alpha\beta}}({\partial_\alpha}{\partial_\beta}\pi-{\Gamma^{\gamma}_{\alpha\beta}}{\partial_\gamma}\pi){g^{\lambda\sigma}}({\partial_\lambda}\pi)({\partial_\sigma}\pi)-{g_{\mu\nu}V(\pi)}\right.\right.\nonumber\\
&&\left.\left.+6\sqrt{6}\alpha({\nabla_\mu}({g^{\alpha\beta}}{{\partial_\alpha}\pi}{{\partial_\beta}\pi}{{\partial_\nu}\pi})-\frac{1}{2}{\nabla_\sigma}({g_{\mu\nu}}{g^{\alpha\beta}}{{\partial_\alpha}\pi}{{\partial_\beta}\pi}{{\partial^\sigma}\pi}))\right\}\right].\label{Huggins}
\end{eqnarray}

In the FLRW background, the energy density and pressure of the DE fluid can be obtained easily from (\ref{eosp}). To obtain the DE Eos in terms of DE scalar field and the geometric quantities we first express $\frac{\rho_{\pi}}{H^2}$ and $\frac{P_{\pi}}{H^2}$ as follows,
\begin{equation}
\frac{\rho_{\pi}}{H^2}=\frac{1}{2}\left(\frac{6{\xi}{\dot\pi}{B'({\pi})}}{{H}(1-6{\xi}B({\pi}))}+\frac{{\dot\pi}^2}{{H^2}(1-6{\xi}B({\pi}))}-\frac{18\sqrt{6}\alpha{\ddot{\pi}{\dot{\pi}}^2}}{{H^2}(1-6{\xi}B({\pi}))}+\frac{2V({\pi})}{{H^2}(1-6{\xi}B({\pi}))}\right)\\ \label{eosp1}
\end{equation}
and
\begin{eqnarray}
\frac{P_{\pi}}{H^2}&=&\frac{1}{2}\left(-\frac{{2\xi}{\ddot{\pi}}{B'({\pi})}}{{{H^2}(1-6{\xi}B({\pi}))}}-\frac{2{\xi}B''(\pi){\dot{\pi}}^2}{{H^2}(1-6{\xi}B({\pi}))}-\frac{4{\xi}{\dot\pi}{B'({\pi})}}{{H}(1-6{\xi}B({\pi}))}\right.\nonumber\\
&&\left.+\frac{{\dot\pi}^2}{{H^2}(1-6{\xi}B({\pi}))}-\frac{6\sqrt{6}\alpha{\ddot{\pi}{\dot{\pi}}^2}}{{H^2}(1-6{\xi}B({\pi}))}-\frac{2V({\pi})}{{H^2}(1-6{\xi}B({\pi}))}\right)\label{eosp2}
\end{eqnarray}
Substituting (\ref{eosp1}) and (\ref{eosp2}) in (\ref{defnomeg}), we get the expression for the DE EoS parameter of our model
\begin{equation}
\omega_\pi=\frac{\left(-\frac{{2\xi}{\ddot{\pi}}{B'({\pi})}}{{{H^2}(1-6{\xi}B({\pi}))}}-\frac{2{\xi}B''(\pi){\dot{\pi}}^2}{{H^2}(1-6{\xi}B({\pi}))}-\frac{4{\xi}{\dot\pi}{B'({\pi})}}{{H}(1-6{\xi}B({\pi}))}+\frac{{\dot\pi}^2}{{H^2}(1-6{\xi}B({\pi}))}-\frac{6\sqrt{6}\alpha{\ddot{\pi}{\dot{\pi}}^2}}{{H^2}(1-6{\xi}B({\pi}))}-\frac{2V({\pi})}{{H^2}(1-6{\xi}B({\pi}))}\right)}{\left(\frac{6{\xi}{\dot\pi}{B'({\pi})}}{{H}(1-6{\xi}B({\pi}))}+\frac{{\dot\pi}^2}{{H^2}(1-6{\xi}B({\pi}))}-\frac{18\sqrt{6}\alpha{\ddot{\pi}{\dot{\pi}}^2}}{{H^2}(1-6{\xi}B({\pi}))}+\frac{2V({\pi})}{{H^2}(1-6{\xi}B({\pi}))}\right)}\label{eospnorm}
\end{equation}
This will be used later in the paper when we shall track the asymptotic time evolution of the EoS parameter. DE being the dominant component of the Universe currently, its EoS parameter determines the fate of the Universe in future. To proceed we first have to track the time evolution of the DE scalar as well as the relevant geometric quantities of the system. Since the system of equations of motion is highly nonlinear we choose to study their asymptotic behaviour using what has come to be known as the dynamical system analysis method in the next section.
\section{Dynamical system analysis} 
In the dynamical system approach, instead of attempting to obtain analytic solutions of the equations of motion one studies the system qualitatively in a phase space made up of dimensionless variables (constructed from the dynamical variables of the system) and their time derivatives. These variables are known as the autonomous variables and their first time derivatives constitute a set of closed system of equations named autonomous system. Depending on the initial conditions the system time-evolves to any of the fixed points of this phase space where the autonomus variables are seen to freeze, thus sealing the fate of the universe described by the original system.
\subsection{Autonomous Equations}
The autonomous equations are a set of first order equations of the type 
\begin{eqnarray}
\frac{{dx}_i}{d\lambda}=f_{i}\left(x_1,x_2,....x_n \right)
\label{auto_sys}
\end{eqnarray}
where $x_{i}\left( \lambda \right)$ are dimensionless variables dependent on some arbitrary parameter $\lambda$ that tracks the evolution of the system and does not appear explicitly on the right hand side.
In an attempt to construct the autonomous system for our model we first define the following autonomous variables  
\begin{eqnarray}
x&=&\frac{{\dot\pi}^2}{{H^2}(1-6{\xi}B({\pi}))}, \qquad y=\frac{2V({\pi})}{{H^2}(1-6{\xi}B({\pi}))}, \qquad  z=\frac{6{\xi}{\dot\pi}{B'({\pi})}}{{H}(1-6{\xi}B({\pi}))},\nonumber\\
\Omega&=&\frac{2\rho_m}{{H^2}(1-6{\xi}B({\pi}))}, \qquad \beta=-{36{\sqrt{6}}{\alpha}H{\dot\pi}}\label{defn}
\end{eqnarray}
so that equation (\ref{f1}) can be rewriten as,
\begin{equation}
{\Omega}+x(1+{\beta})+y+z=1\label{ob}
\end{equation}
which can serve as a constraint on the system. In the above  ${{M_{pl}}^2}=\frac{1}{6}$ is taken  as a convenient choice. We further extend the set $\left\{\Omega, x, y, z, \beta\right\}$ with the following  dimensionless variables
\begin{eqnarray}
{\Sigma} &=& \frac{{B'({\pi})}\pi}{(1-6{\xi}B({\pi}))} \label{Sigma} \\
b &=& \frac{{B''(\pi)}\pi}{B'(\pi)} \label{p1}\\
c &=& -\frac{{V'(\pi)}\pi}{V(\pi)} \label{p2}
\end{eqnarray}
where  ${'}$ denotes derivative w.r.t. $\pi$.
After ascertaining the set of autonomous variables $\left\{x_{i}\right\} \equiv \left\{\Omega, x, y, z, \beta, \Sigma, b, c \right\}$ we construct the set of autonomous equations (\ref{auto_sys}) as
\begin{eqnarray}
{\frac{dx}{dN}} &=&12{\Delta}\frac{x}{z}-2x\left(\frac{{\Theta}}{6}-2\right)+xz,\nonumber\\
{\frac{dy}{dN}} &=&-\frac{cyz}{6 {\xi }{\Sigma}}-2y\left(\frac{{\Theta}}{6}-2\right)+yz,\nonumber\\
{\frac{dz}{dN}} &=&\frac{{z^2}b}{6{\xi}{\Sigma}}+6{\Delta}-z\left(\frac{{\Theta}}{6}-2\right)+{z^2},\nonumber\\
{\frac{d{\Sigma}}{dN}} &=&\frac{z(b+1)}{6{\xi}}+{\Sigma}z,\nonumber\\
{\frac{d\Omega}{dN}} &=&{\Omega}\left[1-\frac{{\Theta}}{3}+z\right],\nonumber\\
{\frac{d\beta}{dN}} &=&{\beta}\left[\left(\frac{{\Theta}}{6}-2\right)+6\frac{{\Delta}}{z}\right]\label{auto}
\end{eqnarray} 
where we choose $N \left( t  \right) = {\rm \ln} \,a\left( t  \right)$ as the independent parameter $\lambda$.  The functions $f_{i}\left(x_1,x_2,....x_n \right)$ on the right hand side are computed using the equations of motion (\ref{f1}), (\ref{f2}) and (\ref{eom}) and the above definition of $N$. Thus the time evolution of the system variables are now tracked with the number of $e$-folding $N$ instead of the comoving time $t$.
We have written equations for variables  $x$, $ y$, $ z$, ${\Sigma}$,  $ \Omega$, $\beta$ and dropped $ \frac{db}{dN}$ and $\frac{dc}{dN}$ to close the autonomous system. Note that with these choices, $b = {\rm constant}$ and $c = {\rm constant}$, $\left(b, c \right)$ become system parameters that characterise the nature of the non-minimal coupling and the potential respectively. From their definitions (\ref{p1}) and  (\ref{p2}) it follows that 
\begin{eqnarray}
B(\pi) &=& {\pi}^{b+1} \label{B} \\
V(\pi) &=& V_0{\pi}^{-c} \label{V}
\end{eqnarray}
Certain points are to be noted in relation to (\ref{auto}). First, while writing equation for $ \Omega$, the equation of continuity has been used. Also two new quantities 
\begin{eqnarray}
\Delta &=& \frac{{\xi}{\ddot{\pi}}{B'({\pi})}}{{{H^2}(1-6{\xi}B({\pi}))}} \label{Deltaold} \\
\Theta &=& \frac{R}{H^2} \label{Theta}
\end{eqnarray}
are defined.These are not independent variables but can be expressed as combinations of the set of autonomous variable and system parameters as,
\begin{eqnarray}
{\Theta}&=&\left[\frac{4x(1+\beta)}{4x(1+\beta)+{(z+{\beta}x)^2}}\right]\left[-6x+12y+9z+\frac{{z^2}b}{2\xi {\Sigma}} +\frac{18(z+{\beta}x)}{(1+\beta)}(-\frac{1}{2}+\frac{zyc}{72{\Sigma}{\xi }x}-\frac{\beta}{12}) \right.\nonumber\\
&&\left.
+3{\lbrace}1-x(1+\beta)-y-z{\rbrace}+3{\beta}x\right],\label{y} \\
{\Delta}&=&{\frac{4xz}{4x(1+\beta)+{(z+{\beta}x)^2}}}\left[-\frac{1}{2}+\frac{zyc}{72{\Sigma}{\xi }x}-\frac{\beta}{12}\right]\nonumber\\
&&-{\frac{z(z+{\beta}x)}{18[4x(1+\beta)+{(z+{\beta}x)^2}]}\left[-6x+12y+9z+\frac{{z^2}b}{2\xi {\Sigma}}+3{\lbrace}1-x(1+\beta)-y-z{\rbrace}+3{\beta}x\right]} \label{Delta}
\end{eqnarray}
Also from the definitions (\ref{defn}), we find that $x$ and $z$ becomes related when $b=1$, resulting in a reduction of the autonomous system. Note that this is a feature of the non-minimal sector entirely irrespective of the presence or absence of the Galileon-like terms in our system (\ref{action}). The special case of $b = 1$ for nonminimally coupled quintissence DE was earlier investigated in \cite{Sami_NMC}. Therefore, in the present paper we confine ourselves to situations where $b \neq 1$. In the next subsection we shall obtain the fixed points of the system (\ref{auto}).
\subsection{Fixed points of the system}
The fixed points $\{x_{\star i}\}$ are defined as points in the phase-space where the autonomous variables stop evolving. Thus, from (\ref{auto_sys}), they are the roots of the system of algebric equations 
\begin{eqnarray}
f_{i}\left(x_1,x_2,....x_n \right) = 0
\label{roots}\end{eqnarray}
For our autonomous system the fixed points are found to be
\begin{flushleft}
\begin{enumerate}
\item $ x_{\star}=1, y_{\star}=0, z_{\star}=0, \Sigma_{\star} \in [-\infty,+\infty],\Omega_{\star} = 0, \beta_{\star} = 0$ \label{1}
\item $ x_{\star} = 0, y_{\star} = 0, z_{\star} = 1, \Sigma_{\star} = -\frac{b+1}{6\xi}, \Omega_{\star} = 0, \beta_{\star} = 0$ \label{2}
\item $ x_{\star} = 0, y_{\star} = 0, z_{\star} = -1+3 \omega, \Sigma_{\star} = -\frac{b+1}{6\xi}, \Omega_{\star} = 2-3 \omega, \beta_{\star} = 0$ \label{3}
\item $ x_{\star} = 0, y_{\star} = \frac{(b+1)(c(1-3\omega)+3(b+1)(1+\omega))}{2c^2}, z_{\star} = \frac{3(b+1)(1+\omega)}{c}, \Sigma_{\star} = -\frac{b+1}{6\xi}, \Omega_{\star} = \frac{2{c^2} - (b+1)(3(\omega+1)(b+c+1)+4c)}{2{c^2}}, \beta_{\star} = 0 $ \label{4}
\item $ x_{\star} = 0, y_{\star} = \frac{5+5b-c}{b+c+1}, z_{\star} = -\frac{2(2+2b-c)}{b+c+1}, \Sigma_{\star} = -\frac{b+1}{6\xi}, \Omega_{\star} = 0, \beta_{\star} = 0 $ \label{5}
\item $ x_{\star} = 0, y_{\star} = 0, z = 0, \Sigma_{\star} \in [-\infty,\infty], \Omega_{\star} = 1, \beta _{\star} \in [-\infty,\infty] $ \label{6}
\item $ x_{\star} = -1, y_{\star} = 0, z_{\star} = 0, \Sigma_{\star} \in [-\infty,\infty], \Omega_{\star} = 0, \beta_{\star} = -2 $ \label{7}
\item $ x_{\star} = 0, y_{\star} = -\frac{(c+5b+5)}{(c-b-1)}, z_{\star} = \frac{(2c+4b+4)}{(c-b-1)}, \Sigma_{\star} = -\frac{(b+1)}{6\xi}, \Omega_{\star} = 0, \beta_{\star} \in [-\infty,\infty] $ \label{8}
\item $x_{\star} = 0, y_{\star} = \frac{(18+3c-c^2)}{(3+c){c^2}}, z_{\star} = -\frac{6}{c}, \Sigma_{\star} = -\frac{1}{3\xi}, \Omega_{\star} = \frac{{c^3}+10{c^2}+15c-18}{(3+c){c^2}}, \beta{\in}[-\infty,\infty]$\label{9}
\end{enumerate}
\end{flushleft}
The fixed points (1) to (5) correspond to $\beta = 0$ (and consequently $\alpha=0$). From the action (\ref{action}), we observe that in this limit our theory of non-minimally coupled Galileon-like DE fulid goes over to a non-minimally coupled quintessence DE which has been discussed exhaustively in \cite{Sami_NMC}. Among the rest of the fixed point (7) corresponds to $\beta \neq 0$ and therefore $\alpha \neq 0$ whereas for the fixed points (6), (8) and (9) $\beta$ is unditermined. These are the fixed points which are non-trivially affected by both to the non-minimal coupling and the Galileon-like term in the scalar field action. Since our purpose is to study whether Galileon-like scalar field with non-minimal coupling (other than derivative coupling) gives any new physics we therefore focus on the fixed points  (6) to (9) in the reminder of the paper.

\section{System dynamics owing to the Galileon-like character}
In the above we have derived the system of autonomous equations(\ref{auto}) and their fixed points following from our model (\ref{action}). From the set of fixed points (1 - 9) we have identified those which bring out the non-trivial effects of the Galilean-like DE fluid on our model, i.e., (6 - 9). In the present section we shall study their ananlytical and numerical features. 

\subsection{Analytical features: Perturbative stability}
Let us start by noting that for the fixed point (\ref{6}) the variables $x, y, z$ which correspond to the the kinetic energy, the potential energy and the NMC of the DE scalar, are all zero. So this is a trivial solution where none of the physically interesting properties of DE change. We therefore discard this one.

We classify the fixed points (7 - 9) according to the stability of the system around them under perturbations $x_{i} - x_{\star i} = \delta x_{i}$. Linearizing the autonomous system with respect to $\delta x_{i}$ we obtain differential equations of the form
\begin{eqnarray}
\frac{d \delta x_{i}}{dN} = {\cal M}_{ij} \delta x_{j}
\label{eigen_value_eqn}
\end{eqnarray}
where ${\cal M}_{ij}$ is a square matrix with elements determined by a given fixed point $x_{\star i}$. A fixed point is stable if all the eigen values $\sigma_{i}$ of the corresponding ${\cal M}_{ij}$ are negative. Otherwise it is unstable.
\begin{enumerate}
\item
The fixed point (\ref{7}) is actually a fixed line as the fixed point value of the variable $\Sigma$ is completely arbitrary. However, everywhere in the autonomous system $\Sigma$ appears along with the variables $z$ whose fixed point value is zero. Thus the eigen values can be determined to be $\sigma_1=-2, \sigma_2=0, \sigma_3=0, \sigma_4=-3, \sigma_5= -1$. Clearly this is an unstable fixed point. 
\item
Next, we consider the fixed point (\ref{8}). Like the previous one this is also a fixed line owing to the complete arbitrariness of $\beta$. But unlike earlier, the eigen-values for the present fixed line can not be computed without reducing it to a fixed point, i.e., fixing $\beta$. We show in the following that this can be achieved by exploiting some physical aspects of our model. 

Using the values $x_{\star i}$ the quantity  ${\Theta}$ is evaluated along the fixed line (8) using equation (\ref{y}) as
\begin{eqnarray}
{\Theta}= \frac{6c(c+5b+5)}{(b+1)(c-b-1)}\label{Y}
\end{eqnarray}
From the definition (\ref{Theta}) ${\Theta}$ is clearly a purely geometric quantity, therefore can be writen explicitely in terms of the scale-factor of the Universe and its derivatives. We integrate that relation useing (\ref{Y}) to chart the asymptotic evolution of scale factor once the Universe freezes into that fixed point. This gives\footnote{A $\star$ in the subscript indicates the variable is calculated on the fixed point or along the fixed line.}
\begin{eqnarray}
a_{\star} \left( t \right)=\frac{a_{0}}{[-m(t-{t_0})]^{\frac{1}{m}}}
\end{eqnarray}
where $a_0$ and $t_0$ are integration constants and $m$ is a combination of the system parameters given by,
\begin{eqnarray}
m=\frac{(c+2b+2)(c+b+1)}{(b+1)(c-b-1)}\label{m}
\end{eqnarray}
Note that the parameters $b$ and $c$ and hence $m$ are determined by the nature of NMC and the self-interaction of the DE scalar. We can also analytically obtain the asymptotic behaviour of $\pi(t)$ at the present fixed line. Let us form the combination of autonomous variables
\begin{equation}
\Gamma = \frac{z}{6 \xi \Sigma} \label{gstat}
\end{equation} 
Along the line (8) it takes the value
\begin{equation}
\Gamma_{\star} = \frac{z_{\star}}{6 \xi \Sigma_{\star}} = -\frac{(2c+4b+4)}{(b+1)(c-b-1)}\label{Gstat} 
\end{equation} 
On the other hand using (\ref{gstat}) with the definition of the autonomous variables in (\ref{defn}) we find 
\begin{eqnarray}
\Gamma = \frac{{\dot{\pi}}}{\pi H} \label{gstat_1}
\end{eqnarray}
which on the present fixed line yields\footnote{Note that the autonomous variables, once they reach the fixed points or lines, stop evolving with $N$. However, the original system variables (e.g., $\pi$) continue to time-evolve. 
}
\begin{eqnarray}
\frac{1}{\pi_{\star}} \frac{d\pi_{\star}}{dt}= \Gamma_{\star} \frac{1}{a_{\star}}\frac{d a_{\star}}{dt} \label{pi}
\end{eqnarray}
Integrating (\ref{pi}), the asymptotic time evolution of $\pi_{\star}$ is found to be 
\begin{eqnarray}
\pi_{\star} \left( t \right)= \pi_{0}\left[ - m \left( t - t_{0} \right) \right]^{-\frac{\Gamma_{\star}}{m}}  =  \pi_{0}\left[ - m \left( t - t_{0} \right) \right]^{\frac{2}{(c+b+1)}}\label{pi_stat}
\end{eqnarray}
where ${\pi}_{0}$ is yet another integration constant. Also in the second equality we have used the definitions (\ref{m}) and (\ref{Gstat}).
Using the asymptotic evolutionary path of the DE scalar $\pi_{\star}\left( t \right)$ and scale factor $a_{\star}\left( t \right)$ in the definition of $\beta$ in (\ref{defn}), we can compute its time-variation along the fixed line (8) as
\begin{eqnarray}
\beta_{\star} \left( t \right) =  - 36 \sqrt{6} \alpha H_{\star} \left( t \right) {\dot{\pi}} _{\star}\left( t \right) =  {(-1)^{\frac{2}{c+b+1}}}{72{\surd{6}}{\alpha}{{\pi}_0}{m^{\frac{1-c-b}{1+c+b}}}}{\frac{(t-{t_0})^{\frac{2}{c+b+1}}}{(t-{t_0})^2}}\label{betaf}
\end{eqnarray} 
From eq. (\ref{betaf}), we see that the time-dependence of $\beta$ can be removed 
if we choose the system parameters $b$ and $c$ such that $ c = - b $. Note that this choice relates the nature of NMC depicted in the function $B \left( \pi \right)$ with the self-interaction $V \left( \pi \right)$ through (\ref{p1}, \ref{p2}) and identifies a class of models where the choice of potential determines the NMC through $B \left( \pi \right) = \int \, d \pi V \left( \pi \right).$

With this choice $\beta$ gets the fixed value 
\begin{eqnarray}
\beta_{\star}=-\frac{72 \sqrt{6} \alpha \pi_{0} (b+2)}{(b+1)(2b+1)} \label{betag}
\end{eqnarray}
and the hitherto fixed line (8) reduces to the fixed point 
\begin{eqnarray}
 x_{\star} = 0, y_{\star} = \frac{(4b+5)}{(2b+1)}, z_{\star} = -\frac{(2b+4)}{(2b+1)}, \Sigma_{\star} = -\frac{(b+1)}{6 \xi},  \Omega_{\star} = 0, \beta_{\star} = -\frac{(b+2)}{(b+1)(2b+1)} \label{fpv}
\end{eqnarray}
Since ${\pi}_0$ is arbitrary in (\ref{betag}), it is chosen to be $ {\pi}_0 = \left( 72 \sqrt{6} {\alpha} \right)^{-1}$ in (\ref{fpv}). Note that this does not remove the effect of the Galileon parameter $\alpha$, signifying the strength of the Galileon term in (\ref{L_Galilean}), from the system but only disguises it within the DE scalar $\pi_{\star} \left( t \right)$ (refer equation (\ref{pi_stat})).

Perturbing the set of autonomous equations about the stationary point (\ref{fpv}) we get the following eigen values:
\begin{eqnarray}
{{\lambda}_1}&=&\frac{2(b+2)(1-b)}{(b+1)(2b+1)},\nonumber\\
{{\lambda}_2}&=&\frac{(P+Q)}{2(b+1)(b+2)(2b+1)},\nonumber\\
{{\lambda}_3}&=&\frac{(P-Q)}{2(b+1)(b+2)(2b+1)},\nonumber\\
{{\lambda}_4}&=&-\frac{(4b+5)(b-1)}{(b+1)(2b+1)},\nonumber\\
{{\lambda}_5}&=&-\frac{(2b+4)}{(2b+1)}
\end{eqnarray}
\noindent Where
\begin{eqnarray}
P&=&-3(10{b^3}+31{b^2}+28b+6)\nonumber\\ Q&=&\surd(324{b^6}+2140{b^5}+6001{b^4}+9040{b^3}
+7652{b^2}+3440b+644) 
\end{eqnarray}
 All these eigen values are negative only when $ b >1 $, so we have stable solutions about the fixed point (\ref{fpv}) for this range of $b-$values. Note that this criteria of finding a stable fixed point further reduces the arbitrariness of our model choice by reducing the completely generic nature of the self-interaction $V \left( \pi \right)$ and therefore also of the NMC $B \left( \pi \right)$.
We thus get a stable vacuum stationary point. Note that,this is a new stable solution which owes its existence to both the NMC and the self-intearction including the Galileon like term in the action (\ref{action}). We shall study our system around this solution in great detail in the next subsection when we take up the numerical analysis.

\item 
(\ref{9}) is again a fixed line owing to the indeterminacy of the fixed point value of the variable $\beta$. 
An analytical approach, as has been charted above, to reduce this line to a fixed point can be taken in this case as well. Alternatively, numerical recipe can also be applied to check its stability. Analysis shows that it is an unstable fixed point.
\end{enumerate}
\subsection{Numerical Analysis}
\begin{figure}[h]
\begin{center}
\includegraphics[width=10cm, height=10cm, angle=0]{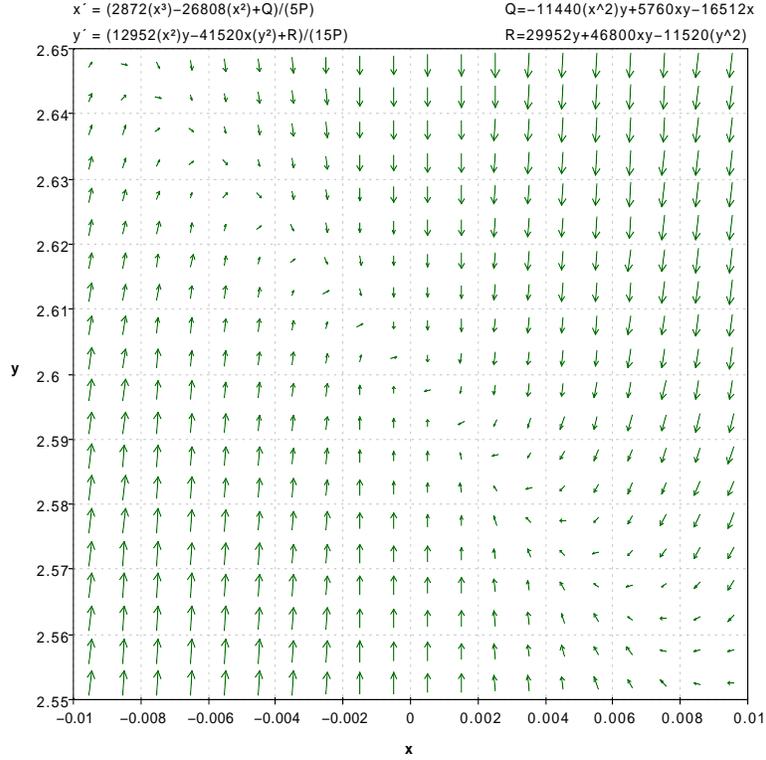}
\end{center}
\caption{phase portrait in the $x-y$ plane for $b=2$ }\label{fp}
\end{figure}

So far, in this section, we have used analytical means to identify, among the set of fixed points in our model, the one that is stable and also owes its existance to both nonminimal nature of the coupling and presence of Galileon-like interaction. This is a novel fixed point hietherto not reported in the literature and we refer to it as the fixed point (\ref{fpv}). Now to get a more detailed picture of the dynamics near this stable fixed point, a numerical study is in order. 
To start with we draw a phase portrait demonstating the existence of the fixed point (\ref{fpv}) under discussion. 
Of course it is not possible to see graphically the whole solutions which lie in the six dimensional hyperspace.  However we generate a phase portrait in the $x-y$ plane, choosing the initial values of the other autonomous variables at their fixed point values, which essentially serves the purpose of displaying the fixed point (\ref{fpv}). 
The figure (\ref{fp}) shows the $x-y$ plane projection of the phase trajectories leading to the fixed point $(0,2.6)$ which is in complete agreement with the values of fixed point (\ref{fpv}) when $b=2$\footnote{Note that the present fixed point (\ref{fpv}) is stable for $b > 1$, so we choose $b  = 2$ throughout this section. }. This approach is equivalent to drawing a two dimensional Poincare section of the numerical trajectory in the six dimensional configuration space. 

A standard $RK~IV$ method was used to numerically solve the set of autonomous dynamical equations, choosing different values of $\xi$ and $\alpha$. Note that these parameters respectively give the proportion of NMC and Galileon term in our model. From the numerical study we have tabulated the futue evolution of the dynamical variables $(x,y,z,\Sigma,\beta,\Omega)$ with $N$, the number of e-foldings, starting from the present epoch $N = 0$. 

\begin{figure}[h]
\hspace{-3.75 cm}
\begin{minipage}[b]{0.40\linewidth}
\centering
\includegraphics[width=\linewidth, angle=-90]{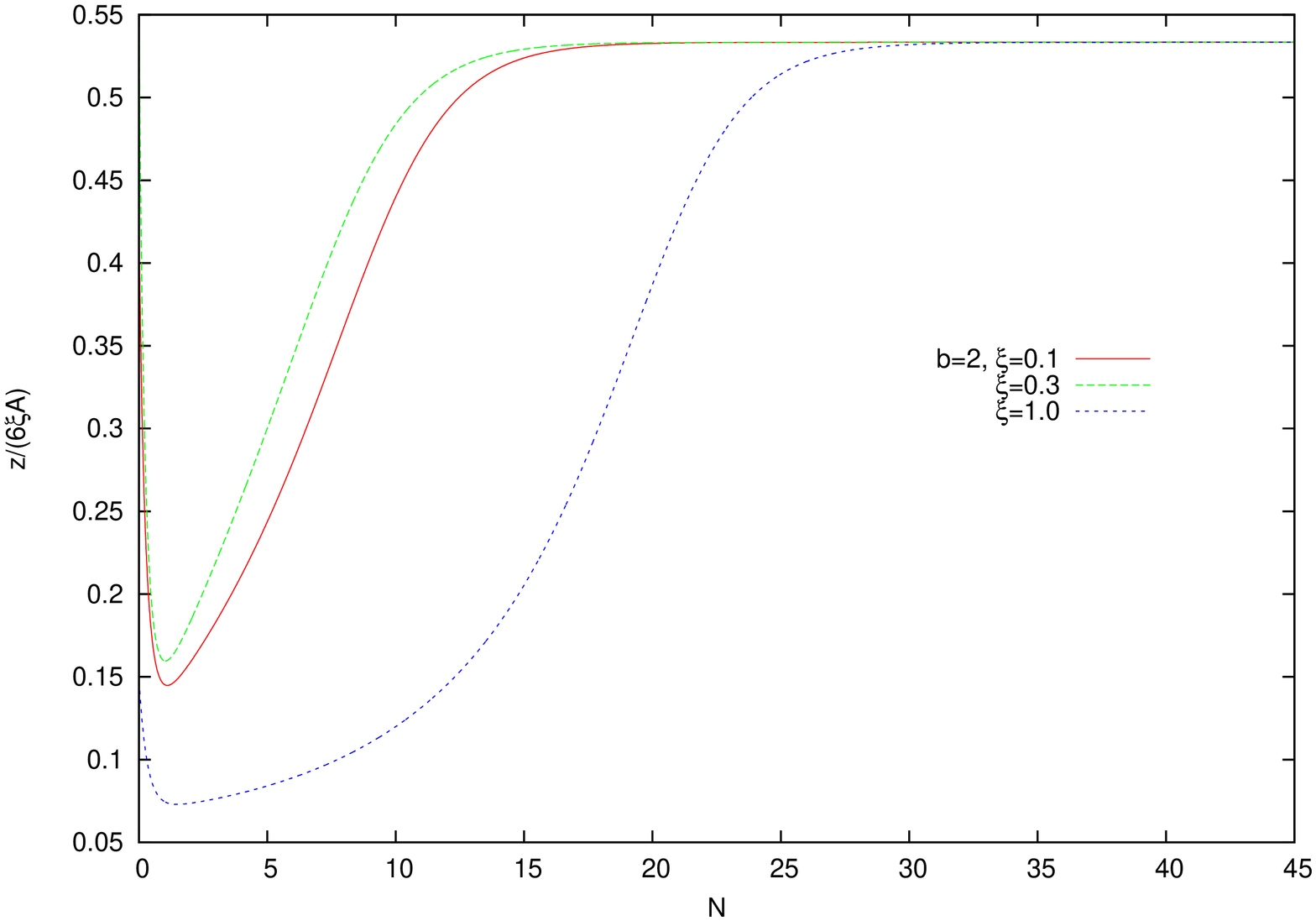}
\caption{The graph for $\frac{z}{6\xi \Sigma}$ .vs. $N$ for different $\xi$ viz. $0.1,0.3,1.0$ where we choose $b=2$.}
\label{rat}
\end{minipage}
\hspace{2.75cm}
\begin{minipage}[b]{0.40\linewidth}
\centering
\includegraphics[width=\linewidth, angle=-90]{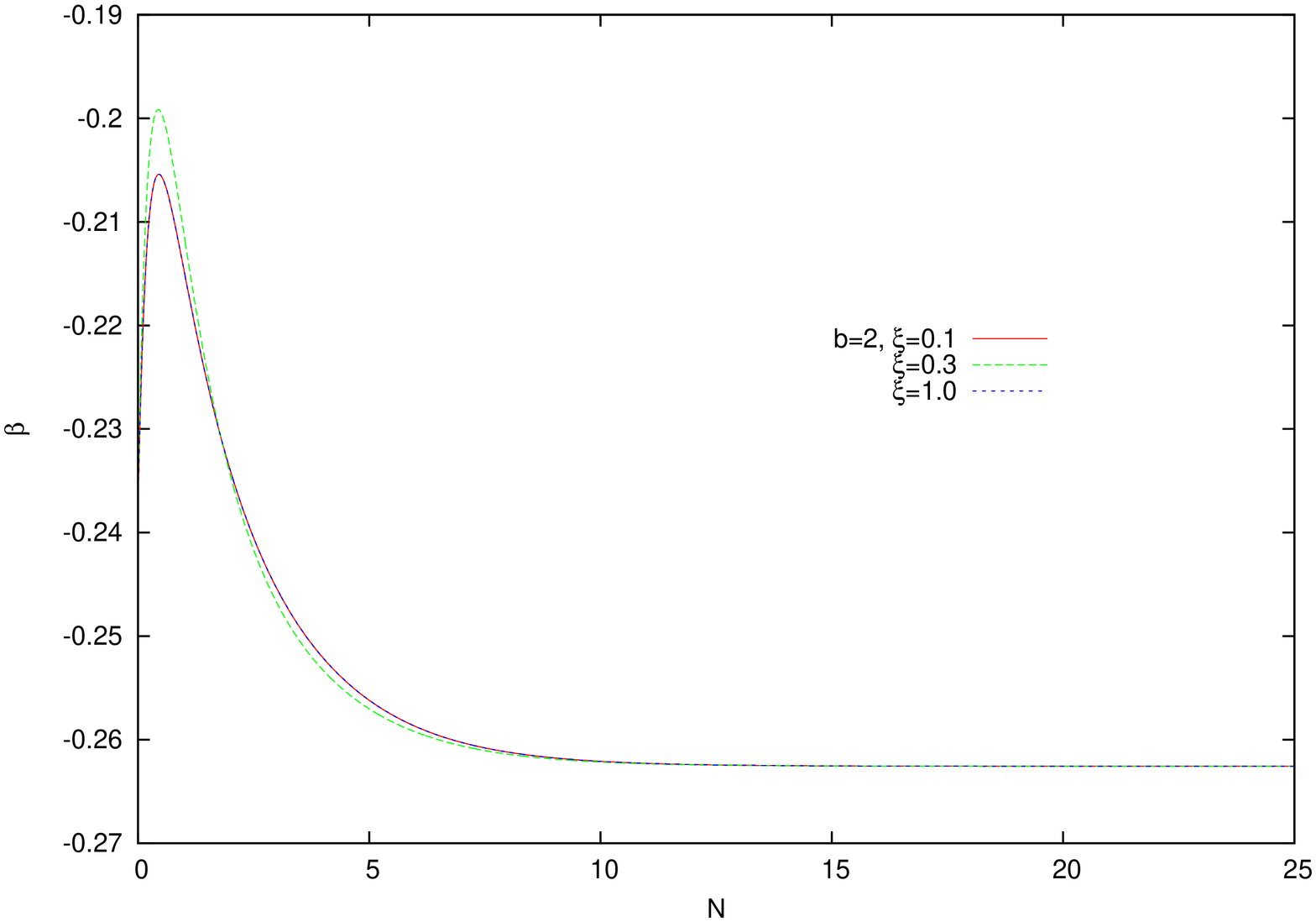}
\caption{The graph for $\beta$ .vs. $N$ for different $\xi$ viz. $0.1,0.3,1.0$ where we choose $b=2$.}
\label{bat}
\end{minipage}
\begin{minipage}[h]{0.50\linewidth}
\centering
\includegraphics[width=\linewidth, angle=-90]{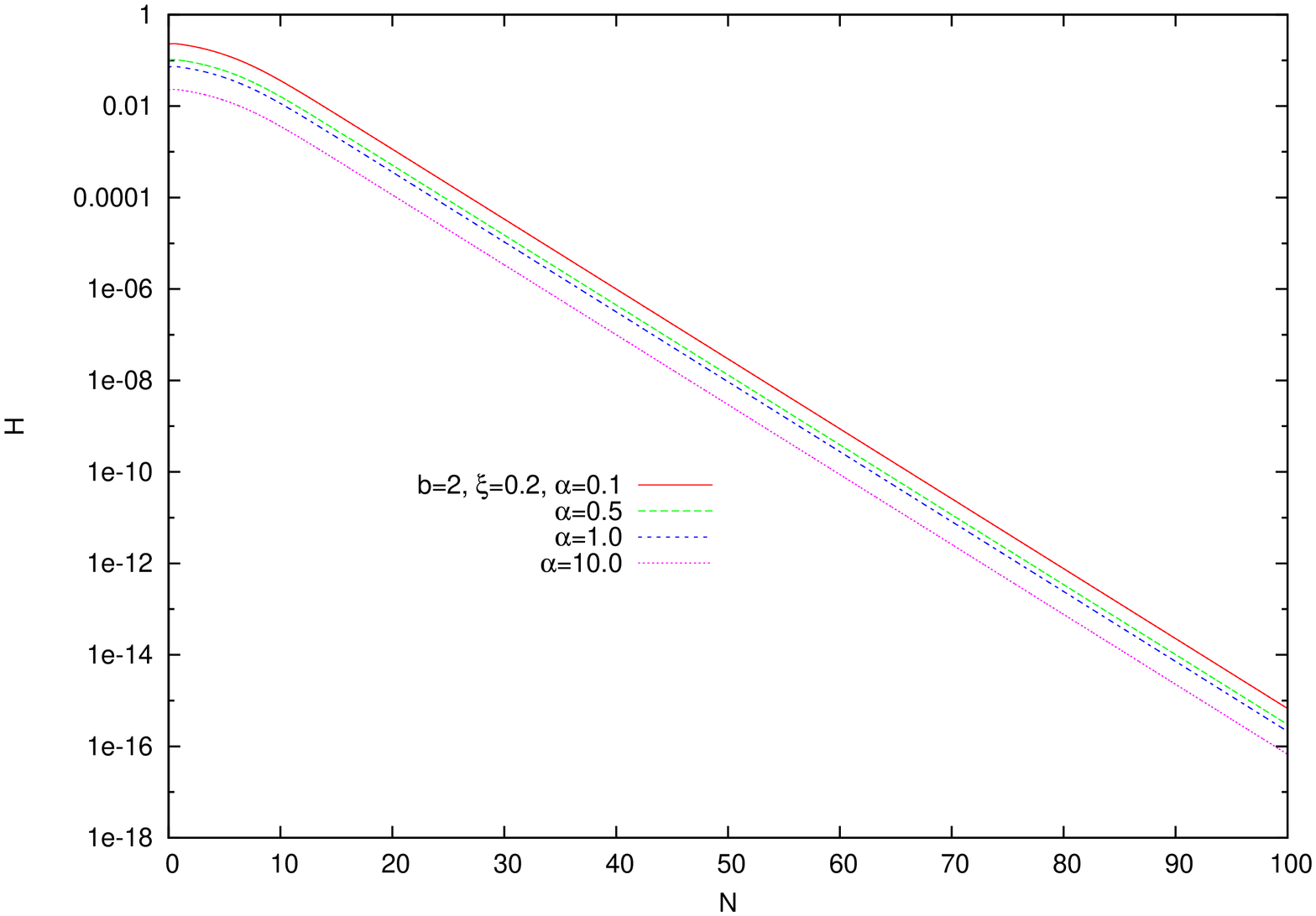}
\caption{A graph of $H$ .vs. $N$ for a fixed $\xi=0.2$ but with widely different $\alpha$ values viz. $0.1,0.5,1.0,10.0$, where $b=2$ is chosen to retain the stability of fixed point.}\label{Hfig}
\end{minipage}
\end{figure}

Using the definitions of the autonomous variables (\ref{auto}) we can express $\Gamma$ defined in equation (\ref{gstat}) as
\begin{equation}
\Gamma = \frac{z}{6\xi\Sigma} = \frac{d\ln\pi}{dN} \label{pi1}
\end{equation}
which can be numerically integrated to find $\pi$ as function of $N$ for the entire dynamics. The equation(\ref{pi1}) will be very useful in the following analysis.
A notable feature of the numerical solution, as depicted in figure \ref{rat}, is that for a given $b-$value, the ratio $\Gamma$, while approaching the fixed point, attains a constant value that remains independent of the choice of $\xi$. Thus the fixed point value of this ratio is independent of the strength of NMC. However the duration of the transient epoch (when the said ratio approaches its fixed point value) is quite sensitive to the choice of $\xi$. From fig(\ref{rat}) one can also identify a characteristic $e$-folding $N_{f}$, a $N-$value beyond which the ratio practically attains its fixed point value. Also note that the dependence of the size of the transient epoch on $\xi$ essentially makes $N_{f}$ $\xi-$dependent as well. This feature enables us to get the analytic expressions for the Hubble function in the asymptotic limit (large $N$). This will be done leter in this section. Also, the autonomous variable $\beta$ is seen to quickly attain a constant value for $N > N_{f}$ in the numerical solution. 

\noindent Using the definition (\ref{defn}), and equations (\ref{gstat}, \ref{gstat_1}) we can show that, 
\begin{equation}
H=\sqrt{\frac{-\Sigma\beta\xi}{6\sqrt{6}{\alpha}z\pi}}\label{NH}
\end{equation}
is an exact relation which is obtained analytically and remains valid troughout the dynamical evolution of the system. 

Using $\pi \left( N \right)$ obtained by numerically integrating (\ref{pi1}) and the numerical solutions of other relevant autonomous variables in the expression of $H$ in equation(\ref{NH}) the behaviour of $H\left( N \right)$ has been charted out  in fig.(\ref{Hfig}) for different values of $\alpha$ ranging from $0.1-10.0$ and a given of $\xi=0.2$. Numerically, we observe that $H$ decreases exponentially with $N$ and there is an $\alpha$ dependence as well. 

To depict this exponential behaviour, the $H$-axis is set in logarithmic scale in fig.(\ref{Hfig}). Remember that $\alpha$ measures the strength of the Galileon character of the DE scalar field. So we find that the Galileon like term in the scalar field action has non-trivial effect on the Hubble parameter. A detailed discussion on the functional form in the $N > N_{f}$ regime will be given later in this section. 
For now, let us note that since the ratio $\Gamma$ and the autonomous variable $\beta$ both are seen to attain a constant value in the numerical integration beyond $N = N_{f}$, for a given value of $\xi$, (see fig(\ref{rat}) and fig(\ref{bat})), equation (\ref{NH}) immediately tells us that the behaviour of the Hubble function $H$ is dominated by the behaviour of $\pi(N)$ in this region. We also see that $H$ scales as $\alpha^{-\frac{1}{2}}$.

\begin{figure}[h]
\hspace{-3.75 cm}
\begin{minipage}[b]{0.40\linewidth}
\centering
\includegraphics[width=\linewidth, angle=-90]{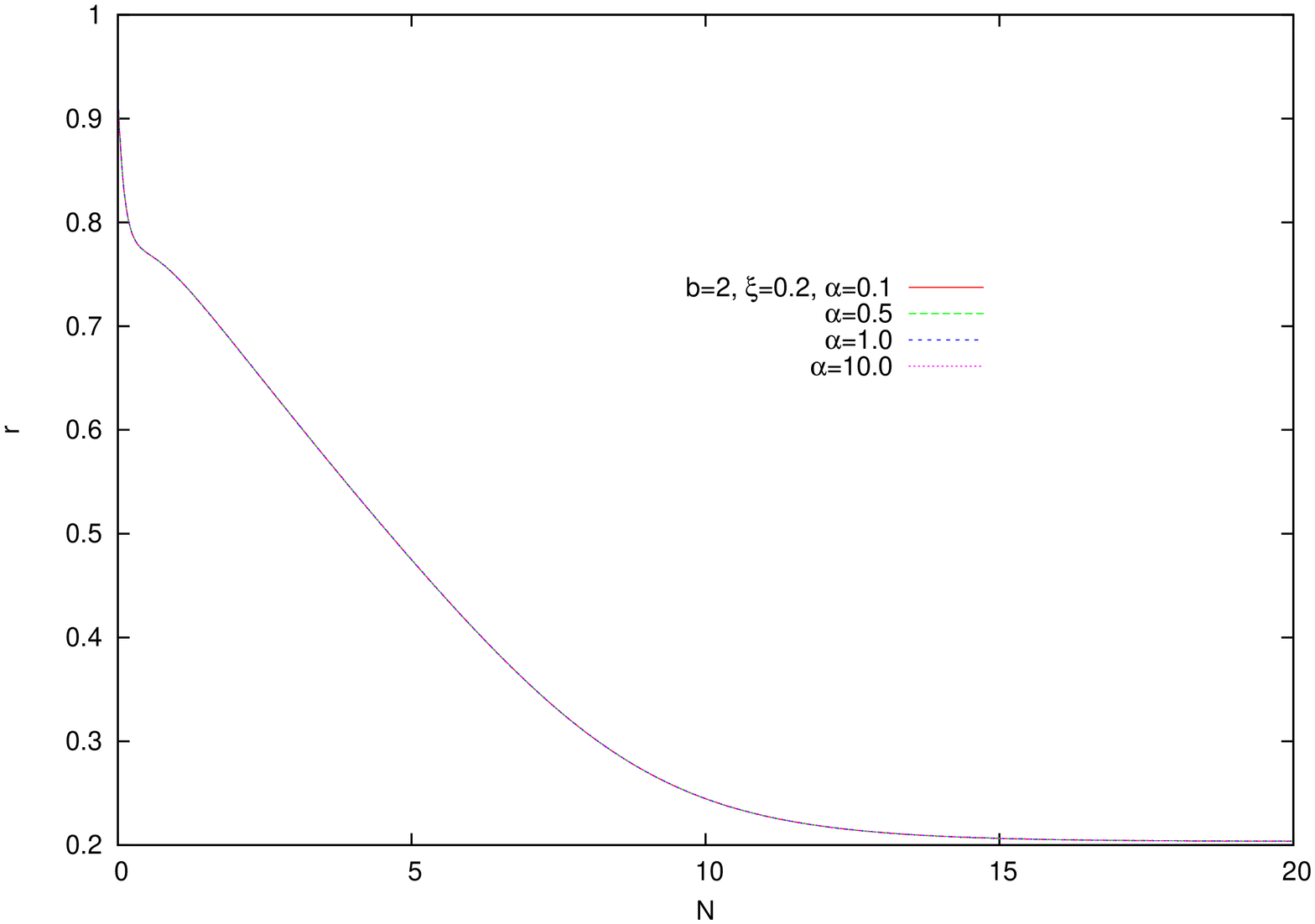}
\caption{A graph showing $r(N)$ vs $N$ for $\xi=0.2$, with varying $\alpha$ values for two decades.}\label{ra}
\end{minipage}
\hspace{2.75cm}
\begin{minipage}[b]{0.40\linewidth}
\centering
\includegraphics[width=\linewidth, angle=-90]{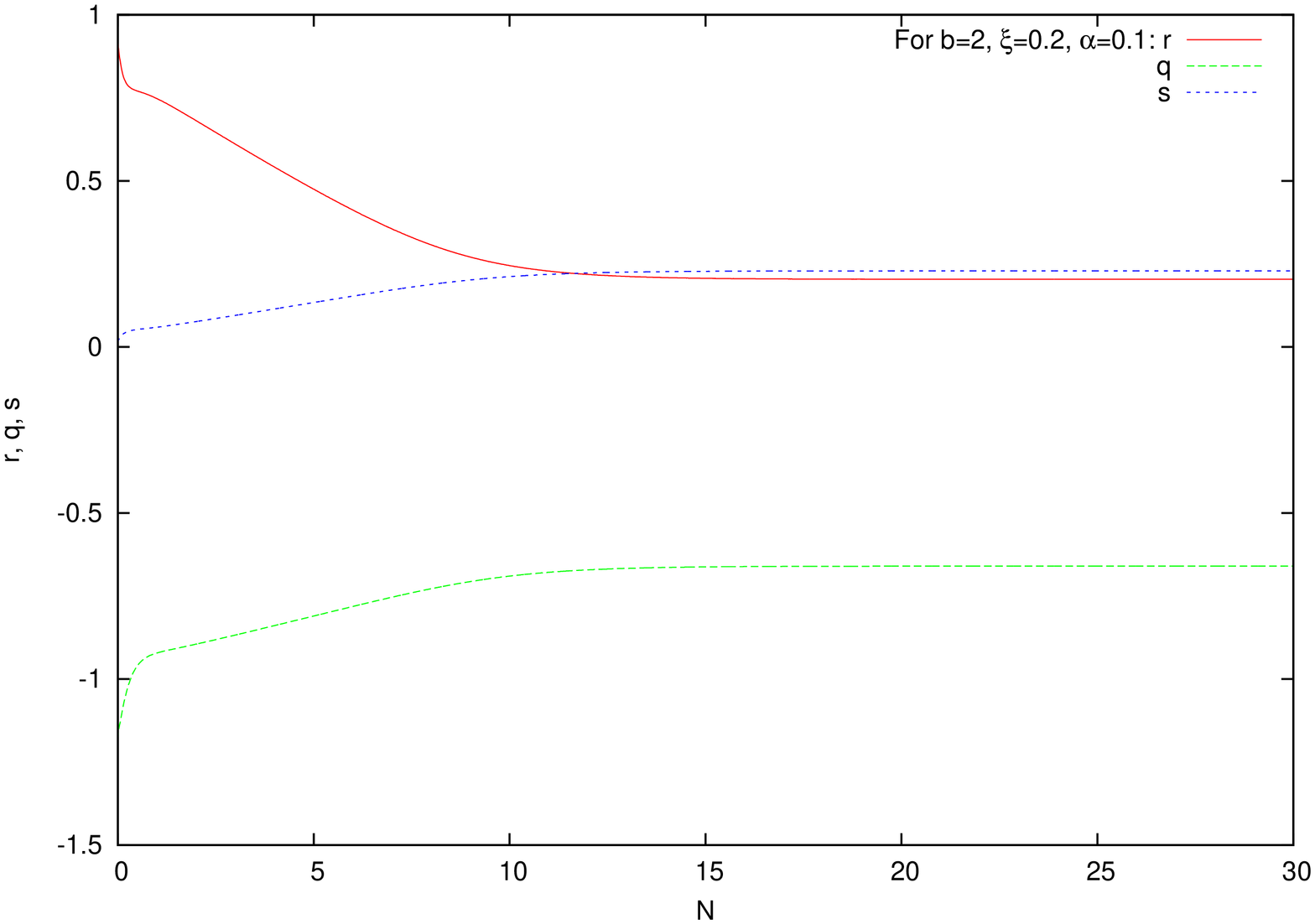}
\caption{A graph showing $r(N)$, $q(N)$ and $s(N)$ for $\xi=0.2$, $\alpha=0.1$ and $b=2$ }\label{rqs}
\end{minipage}
\end{figure}
To compare results of our model with the standard model we use the versatile $r-s$ diagram \cite{vsahni}. Here along with the deceleration parameter $q$ two new parameters $r$ and $s$ are defined as
\begin{eqnarray}
q = -\frac{\ddot{a}a}{\dot{a}^2}, \qquad r = \frac{{\dddot{a}}{a^2}}{{\dot{a}}^3}, \qquad s = \frac{r-1}{3(q-\frac{1}{2})}
\label{rs}
\end{eqnarray}
These parameters can be readily expressed in terms of the Hubble parameter and its time derivatives as
\begin{eqnarray}
q =  -1 - \frac{\dot{H}}{H^2}, \qquad  r = \frac{H^3+3H\dot{H}+\ddot{H}}{H^3}, \qquad s = \frac{r-1}{3(q-\frac{1}{2})}
\label{rs11}
\end{eqnarray}
and hence they depend only on the geometric properties of any cosmological model. Thus in a $r-s$ plane different cosmological models follow different trajectories or occupy different locations. 

Though $H(N)$ varies with $\alpha$ when $\xi$ is kept fixed, a plot of $r(N)$ (see fig.(\ref{ra})) for widely different $\alpha-$values ($\alpha = 0.1 - 10.0$) and fixed $\xi$, ($\xi=0.2$) shows that the behaviour of $r(N)$ is independent of the choice of $\alpha$ when $\xi$ is kept fixed. Similar behaviour is found for $q$ and $s$ as well. This can be explained easily by noting that the pair $\left(q, r \right)$ defined in (\ref{rs11}) can be equivalently represented by the Hubble parameter and its derivatives with respect to the $e-$folding as
\begin{eqnarray}
q =  -1 -  \frac{1}{H}\frac{dH}{dN}, \qquad r = 1 + \frac{3 }{H} \frac{d H}{d N} + \left( \frac{1}{H} \frac{dH}{dN}\right)^{2}+ \frac{1}{H}\frac{d^{2}H}{d N^{2}}, 
\label{rs2}
\end{eqnarray}
which shows that the $\alpha-$dependence in $H$, as has been discussed bellow equation (\ref{NH}), will get cancelled. Since $s$ is constructed out of $\left(q, r \right)$ it also behaves similarly.
So in fig.(\ref{rqs}), we show the behaviour of the trio $\left(q\left(N\right), r \left( N \right), s \left( N \right)\right)$ obtained from our numerical study for $\xi=0.2$ and a single $\alpha-$value $\alpha=0.1$. Refering to figure{\ref{rat}} one may readily observe that all the three geometric parameters saturate asymptotically beyond $N = N_{f}$.  

\begin{figure}[h]
\hspace{-3.75 cm}
\begin{minipage}[b]{0.40\linewidth}
\centering
\includegraphics[width=\linewidth, angle=-90]{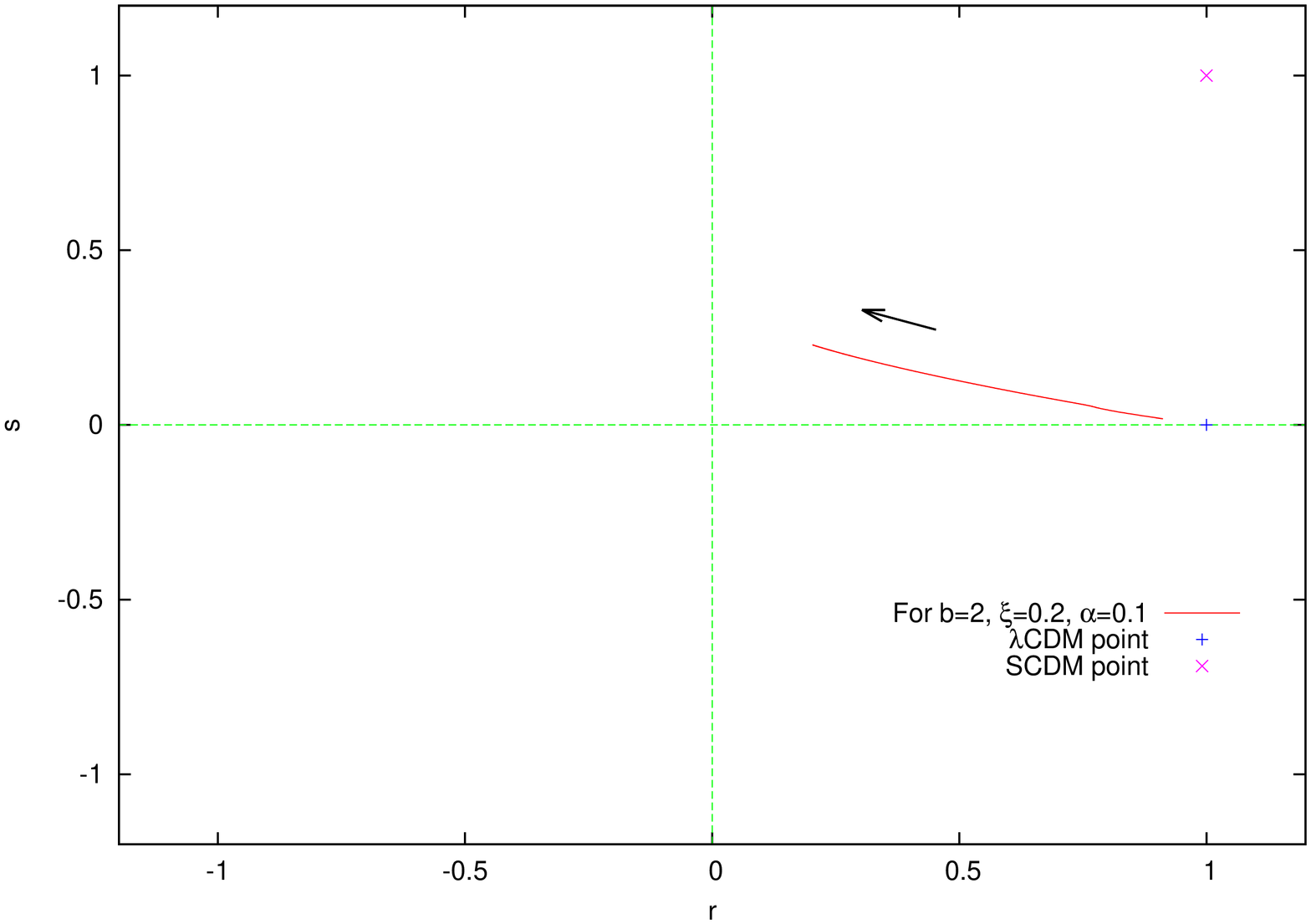}
\caption{The $r-s$ diagram obtained from the data of fig.\ref{rqs}. The arrow shows the direction of evolution.}\label{rs1}
\end{minipage}
\hspace{2.75cm}
\begin{minipage}[b]{0.40\linewidth}
\centering
\includegraphics[width=\linewidth, angle=-90]{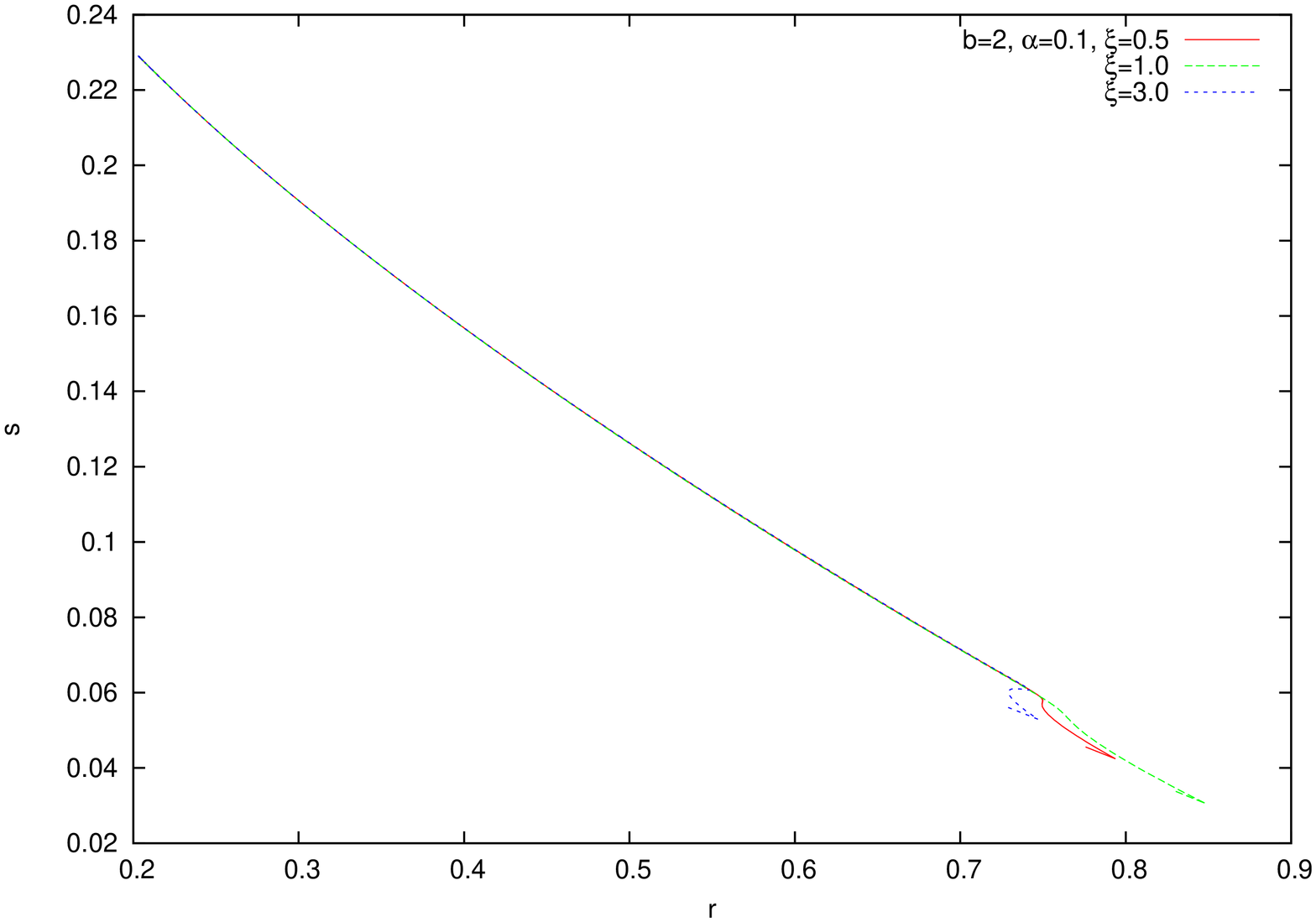}
\caption{The $r-s$ plot obtained for different $\xi-$values and $\alpha = 0.1$ in the transient epoch.}\label{rsxi}
\end{minipage}
\end{figure}

In figure (\ref{rs1}) we use the same data of fig(\ref{rqs}) to plot the corresponding $r-s$ diagram. It is found that the our model is moving away from the $\Lambda$CDM point for large $N$ and finally settles around $(0.202,0.23)$, as all the three parameters $(r, q, s)$ attain a steady value (see fig. \ref{rqs})
. The overall nature of $r-s$ plot in this dynamics found from our model is roughly a straight line. It is found that the variation of $r(N)$, $q(N)$,$s(N)$ for different $\xi$ values do not show any significant change. Shifting our focus to the transient epoch (i.e. $N<N_f$) we see the $\xi$ dependence in the $r-s$ plot (see fig. \ref{rsxi}) which has been mentioned earlier in context of fig. \ref{rat}. However, beyond the transient epoch all the plots for different $\xi-$values overlap on each other near the fixed point under discussion. 

As proposed earlier, we now provide some analytic estimations in the asymptotic regime. To be specific, we refer to the epochs where $N > N_{f}$ as the asymptotic region. Here, the ratio $\Gamma$ in the left hand side of eq. (\ref{pi1}) takes some constant value, say $k$, as mentioned before and we can readily integrate the relation to obtain
\begin{eqnarray}
\pi(N) &=& \pi(N_f)~e^{k(N-N_f)} = \pi_f ~e^{k(N-N_f)}
\label{thpi}
\end{eqnarray}
where $\pi_{f}$ is the value of the $\pi$ field at $N =N_{f}$.
Now using eq. (\ref{thpi}) to substituting for $\pi$ in eq. (\ref{NH}), one may obtain a closed form 
expression for the $H(N)$ in the asymptotic region as
\begin{equation}
H
= H_f~e^{-\frac{k}{2}(N-N_f)}
\label{thNH}
\end{equation}
with $H_f=\sqrt{\frac{-\Sigma\beta\xi}{6\sqrt{6}{\alpha}z \pi_f}}$. Note that since $\beta$ takes negative values(\ref{fpv}) near the fixed point so $H$ is always real. Further, $H_f$
is independent on $N$, but interestingly varies with$ \sim \frac{1}{\sqrt{\alpha}}$. This prototype dependence of $H(N)$ on $\alpha$ in the asymptotic regime was referred earlier in connection to the fig(\ref{Hfig}) though this exact functional dependence has not been claimed from the numerical data. 

Armed with the numerical solution of our system near the fixed point (\ref{fpv}) we now proceed to calculate the corresponding equation of state parameter in the next subsection.
\subsection{The equation of state parameter in terms of autonomous variables}
\begin{figure}[h]
\begin{center}
\includegraphics[width=10cm, height=10cm, angle=-90]{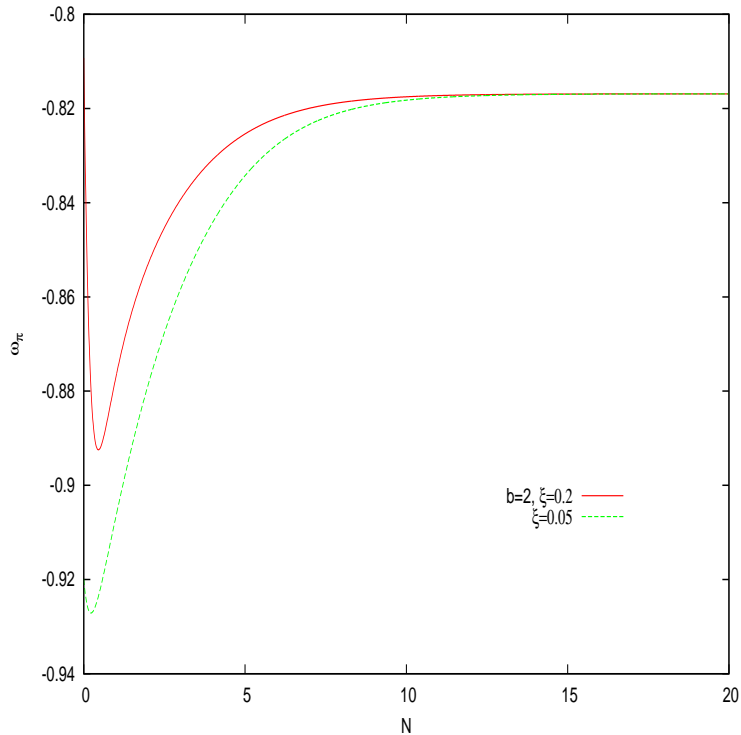}
\end{center}
\caption{A graph of $\omega_{\pi}$ vs $N$ for different values of $\xi$, where $b=2$ is chosen to retain the stability of fixed point.}\label{omega_pi}
\end{figure}

As we have mentioned in the introduction NMC has a direct influence on the energy momentum tensor and through it on the equation of state (see eq. (\ref{defnomeg})) parameter. To use the numerical solutions to work out the evolution of the EoS parameter we first have to re-expresse equations (\ref{eosp1}),(\ref{eosp2})) in terms of the dimensionless autonomous variables as,
\begin{eqnarray}
\frac{\rho}{H^2} & = & \frac{1}{2}(z+x+\frac{3x\Delta\beta}{z}+y) \label{eospp1} \\
\frac{P}{H^2} & = & \frac{1}{2}(-2\Delta-\frac{bz^2}{18\Sigma\xi}-\frac{2z}{3}+x+\frac{x\Delta\beta}{z}-y) \label{eospp2} \\
\end{eqnarray}
so we can rewrite the equation-((\ref{eospnorm}) as
\begin{eqnarray}
\omega_\pi=\frac{(-2\Delta-\frac{bz^2}{18\Sigma\xi}-\frac{2z}{3}+x+\frac{x\Delta\beta}{z}-y)}{(z+x+\frac{3x\Delta\beta}{z}+y)}\label{eospauto}
\end{eqnarray}
Using the numerical solution for the system near the present fixed point we plot the EoS parameter $\omega_{\pi}$ for DE against the number of e-foldings ($N$) in fig. (\ref{omega_pi}). The plot shows that the value of  $\omega_{\pi}$ saturates to $< -\frac{1}{3}$ at high value of $N$ for a range of values of the NMC parameter $\xi$. However, it does not cross over to the phantom regime $< -1$ for any $N-$value, although can approach the phantom threshold quite closely near the present epoch.
As the EOS parameter for dark energy $\omega_{\pi}$ is found to lie below -$\frac{1}{3}$ for large values of $N$, it indicates that the universe will remain in an accelerating phase. 

\section{Conclusion}
We have considered a scalar field with non-minimal coupling (NMC) and power-law potential. The scalar field action contains, apart from the usual canonical kinetic energy, a cubic kinetic energy term like $\Box{\pi} (\nabla \pi)^{2}$ which is reminiscent of Galileon scalar models, therefore we call the scalar of our model Galileon-like. This nonlinear (higher derivative) term allows the Vainshtein mechanism \cite{AASen12} to kick in, though keeps the equation of motion second order to avoiding the Ostrogradsky ghosts, like a characteristic Galileon scalar does. We also included a generic self-interaction potential to avoid the Galileon shift symmetry, since it has been shown that any non-minimally coupled DE scalar field with non-standard (non-linear) kinetic terms and Galilean shift-symmetry (like the Galileon field) is ruled out due to too large a variation in the effective Newton's constant \cite{solar_system_rule_out}. Coupling constants were introduced to control the strength of the NMC and the Galileon interaction. 

The NMC has a specific appeal in relation to the requirements of dark energy fluids. With minimal coupling the equation of state parameter (EOS) for Quintessence is confined within the range $-\frac{1}{3}<\omega<-1$. But observations suggest that the EOS can be less than $-1$. It has been shown that non-minimally coupled Quintessence model can cross $\omega=-1$.  Thus, with NMC,  one could get into ``Phantom" regime without the introduction of non-canonical kinetic energy (Phantom) term. In this paper we have investigated if a generically non-minimally coupled Galileon-like Dark Energy (DE) model shares this feature. The focus of the present paper was thus on the possible effect of introducing a term with Galileon character within the non-minimally coupled Quintessence model. We have found that the dynamics of space-time geometry is definitely influenced.  By analyzing the autonomous equations we have shown that a new stable fixed point owing its existance to the cubic Galileon term in the scalar field action arises for certain choice of the NMC and self-interaction potential. Taking appropriate values for the corresponding coupling constants we have seen that our results agree with \cite{Sami_NMC} which serves as a consistency check. 

From a numerical analysis of the dynamics at the vicinity of the new stable fixed point, influence of Galileon like term was found more in a transient phase rather than in the asymptotic phase. To compare our model with the standard cosmological models we have utilized the celebrated state finder technique \cite{vsahni}. Finally the numerical solution was used to find the the dynamical EOS parameter (see fig \ref{omega_pi}). It is observed that initially the model moves closer to the Phantom threshold near the present epoch before finally going over to the non Phantom accelerated regime asymptotically. Inclusion of a cubic Galileon-like kinetic energy term in the non-minimally coupled quintessance model is thus demonstrated to be of considerable interest. It revealed the existence of a new channel where the universe exhibits the accelerated evolution and resides al along within the non-phantom regime. Our result thus extends the findings of \cite{Cubic_Galileon_NMC} for more general NMC than linear and quadratic couplings.

\section*{Acknowledgement}
The authors thank the referee for useful comments. SNB acknowledges the support by UGC Minor Research Grant File No. PSW­162/14­15(ERO). AS acknowledges the support by DST SERB under Grant No. SR/FTP/PS-208/2012.


\end{document}